\documentclass[%
 reprint,
nofootinbib,
 amsmath,amssymb,
 aps, amsfonts,
prd,
floatfix
]{revtex4-2}
\usepackage{microtype}

\usepackage{hyperref}
\hypersetup{
    colorlinks=true,
    linkcolor=[rgb]{0,0,0.6},      
    urlcolor=[rgb]{0,0,0.6},
    citecolor=[rgb]{0,0,0.6}
    }

\usepackage[normalem]{ulem}
\usepackage{soul}
\setcitestyle{numbers,open={[},close={]}}
\usepackage{graphicx}
\usepackage{dcolumn}
\usepackage{bm}

\usepackage[super]{nth}
\usepackage{booktabs}
\usepackage[version=4]{mhchem}

\usepackage{Zallman,lettrine}

\newcommand{\dd}{\mathrm{d}}

\newcommand{\bigO}{\mathcal{O}}

\newcommand{\orcid}[1]{\href{https://orcid.org/#1}{\includegraphics[width=10pt]{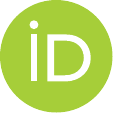}}}

\begin{document}

\preprint{APS/123-QED}

\title{Dark Matter and General Relativistic Instability in Supermassive Stars}

\author{Kyle S. Kehrer\,\orcid{0000-0002-8714-1599}}
\email{kkehrer@ucsd.edu}
\affiliation{Department of Physics, University of California San Diego, La Jolla, California 92093, USA}
\author{George M. Fuller\,\orcid{0000-0002-4203-4108}}%
\email{gfuller@physics.ucsd.edu}
\affiliation{Department of Physics, University of California San Diego, La Jolla, California 92093, USA}

\date{\today}

\begin{abstract}
We calculate the extent to which collisionless dark matter impacts the stability of supermassive stars $(M\gtrsim10^4\,M_\odot)$. We find that, depending on the star's mass, a dark matter content in excess of ${\sim}1\%$ by mass throughout the entire star can raise the critical central density for the onset of general relativistic instability, in some cases by orders of magnitude. We consider implications of this effect for the onset of nuclear burning and significant neutrino energy losses. 
\end{abstract}

\maketitle


\section{Introduction}

In this paper, we examine how dark matter (DM) could impact the onset of general relativistic instability in massive self-gravitating configurations, i.e., supermassive stars (SMSs), as well as their nuclear and neutrino evolution. SMSs are supported principally by radiation pressure, making them vulnerable to general relativistic instability. If these extreme objects had existed in the early universe, and then collapsed to black holes, they could provide massive seeds that facilitate early production of supermassive black holes (SMBHs).

Indeed, SMBHs seem to be extant at an embarrassingly early epoch in the history of the universe. For example, recent high redshift surveys from the James Webb Space Telescope (JWST)~\cite{jwstsmbh} and even the Chandra X-Ray Observatory~\cite{chandrasmbh} suggest the existence of compact objects with masses ${\gtrsim}10^7\,M_\odot$ at redshift $z\sim 10$, when the universe is a scant ${\sim}400\,{\rm Myr}$ in age. These observations are consistent with lower redshift observations (e.g.\ \cite{Mortlock_2011,Venemans_2013,Wu_2015,Ba_ados_2017,Venemans_2017,2023ApJ...953L..29L}) that also suggest a significant inventory of SMBHs early on. Moreover, these black holes may be growing too quickly to be consistent with building up the observationally-inferred SMBH masses given a ${\sim}1\text{--}10\,M_\odot$ progenitor black hole, as expected in the standard pictures of stellar evolution and mass accretion. 

By contrast, at the end of their lives, SMSs would likely collapse directly to black holes, providing large SMBH ``seeds'' that would be able to grow to the SMBH sizes observed in the early universe. Such scenarios are favored by some modelers since they do not invoke super-Eddington accretion rates to explain the high masses observed~\cite{2007MNRAS.377L..64L,2006MNRAS.371.1813L}. \citet{chandrasmbh} reports having observational evidence that supports this argument. The collapse of the SMS itself is potentially observable with future gravitational wave observatories like DECIGO~\cite{decigo}. This gravitational wave signal stems, in part, from an assumed (slightly) non-spherically symmetric neutrino burst generated from the collapsing homologous core~\cite{lifuller}. 

For the purposes of this paper, an ``SMS'' is any hydrostatic and self-gravitating gas cloud with a mass in excess of ${\sim} 10^4$ $M_\odot$. These are fully convective, high entropy systems whose pressure support is almost entirely derived from radiation (i.e.\ relativistic particles). They will eventually suffer general relativistic instability~\cite{fh1962,iben1963,fowler1964,fww86}. Such configurations, those that in general rely on pressure from relativistic particles, are ``trembling on the verge of instability\footnote{Phrase attributed to W.A.\ Fowler.},'' and are most widely realized in compact scenarios like white dwarfs and neutron stars. In Newtonian gravitation, radiation dominated systems are nearly neutrally buoyant under radial perturbations. If the pressure is entirely derived from radiation, the outward pressure forces always perfectly match the inward gravitational forces as the object is expanded and contracted, meaning such perturbations cost no energy to perform, i.e., a metastable configuration.

The situation changes, however, when first order general relativistic (GR) corrections are included in the analysis. Being non-linear, GR predicts that both stress-energy (primarily baryon rest mass, in this case) \emph{and spacetime curvature itself} cause local spacetime curvature. This makes the metastable system under Newtonian gravity actually marginally unstable under GR. This is the Feynman-Chandrasekhar post-Newtonian instability~\cite{fc,Feynman:1996kb}.

Specifically, upon the star's core exceeding a critical central density, it will become unstable to collapse and unable to establish hydrostatic equilibrium. It will most likely collapse directly into a black hole, but may explode completely if energy derived from nuclear reactions (or some other source) can be added quickly enough to make up for the in-fall kinetic energy of collapse and neutrino energy losses \cite{1973ApJ...183..941F}. If the energy source in this scenario indeed stems from nuclear reactions, then an explosion is only possible if the star has sufficient metallicity to burn Hydrogen with the more efficient Carbon-Nitrogen-Oxygen (CNO) cycle instead of through the Weak Interaction limited proton-proton (PPI) chain~\cite{fww86}.

While the formation of such SMSs seems dubious at first, especially when considering cloud fragmentation, \citet{jwst} argues that tantalizing evidence of such objects may have been observed by JWST. This, however, begs the question as to how they would have been observed in the first place since they are so close to instability. \citet{dsreview} puts forward the idea that these SMSs could be supported through particle DM self-annihilation into standard model particles which would keep them ``puffy'' (i.e.\ lower the central density) and prevent GR instability, at least until they exhaust their DM reserves. 

Alternatively, a SMS also could be stabilized by DM without assuming such a self-annihilation cross section~\cite{mclf,Bisnovatyi-Kogan_1998}. If it is assumed that the DM particles are collisionless and only interact with the Standard Model component of the star through gravitation, then it is possible to derive how much its presence alone raises the critical central density beyond which the star suffers the Feynman-Chandrasekhar instability.

Non-intuitively, adding more matter to the meta-stable star, specifically a non-interacting non-relativistic fluid, actually tends to stabilize the configuration as opposed to tipping it closer to collapse. The physical mechanism of this increased stability arises from the DM acting as a source of non-relativistic stress.

Absent a significant DM content, heavier (${\gtrsim}10^5\,M_\odot$) SMSs will undergo the Feynman-Chandrasekhar instability prior to, or just after Hydrogen burning ignition, meaning they effectively have no main sequence phase (see e.g.~\citet{fww86}). An interesting question is then whether a sufficient DM content in these objects could delay the onset of instability, allowing a hydrostatic burning phase, and thereby changing the downstream nuclear and entropy evolution of these stars.

It will be shown in what follows that the amount of DM needed to substantially change where this instability occurs is rather extreme. For example, the DM density at the galactic center or in DM halos considered in standard cosmology are small compared to the those considered here. However, the early universe and some proposed DM candidates, e.g. self interacting models, may provide the necessary DM densities in the compact configurations considered below. It has been argued that the DM density can be greatly increased via adiabatic contraction, a purely gravitational process. For example, this has been proposed in works on ``Dark Stars,'' e.g.~in Ref.~\cite{1986ApJ...301...27B}. We return to these issues in the conclusion of this paper.

In Section \ref{sec:stability}, we establish the overall framework to analyze stability of SMSs by considering the case without any DM. We determine the critical central density from extremizing the total energy of the configuration with a \nth{1} Order GR correction term to derive the standard result. 

In Section \ref{sec:DM}, we examine two kinematic limits of the DM based on its specific kinetic energy relative to the gravitational potential at the SMS's surface, determine the total energy added by this DM, and derive how its presence impacts the critical central density as a function of the DM mass fraction. We then present our results for several SMS masses to examine how the effect also depends on the total mass. 

In Section \ref{sec:response}, we approximate how the total fraction of DM changes with radial perturbations of the SMS, providing insight into the physical mechanism for the enhanced stability discussed in the previous section. 

In Section \ref{sec:laneemdenf}, we determine the extent to which the equilibrium SMS density distribution is impacted by explicitly considering the gravitational influence of the DM. This analysis is performed for several SMS masses and DM mass fractions.

Finally, in Section \ref{sec:ppi}, we investigate nuclear burning rates, neutrino emission rates, and their respective luminosities as a fraction of the stars' Eddington luminosities. This allows us to ascertain how ``important'' nuclear burning is to an SMS's evolution and when neutrino emission overpowers it. We also determine by how much these luminosities change with the addition of DM, and what that may mean for the downstream stability and entropy evolution. 

\section{Stability of Supermassive Stars}\label{sec:stability}

The analysis in this section will closely follow~\citet{mnras}'s analytical treatment, specifically their ``Approach II'' in Section 2.5, to derive the onset of stability for SMSs. The SMSs of concern in this paper are taken to be spherically symmetric, with primordial metallicity, fully ionized and fully convective. They are high entropy gas clouds supported against gravity almost entirely by radiation pressure $P_\text{rad.}$, with a marginal but important contribution from classical Boltzmann gas pressure $P_\text{gas}$. We define the parameter $\beta$ as the ratio of gas pressure to total pressure (the final approximation following from~\citet{g1})
\begin{equation}\label{eq:beta}
    \beta = \frac{P_\text{gas}}{P_\text{gas}+P_\text{rad.}} \approx \frac{4.3}{\mu}\left(\frac{M}{M_\odot}\right)^{-1/2},
\end{equation}
where $\mu=0.59$ is the mean molecular weight (in atomic mass units) per particle for a primordial composition gas. 

Given our approximations, the equation of state (EOS) for this configuration can be modeled accurately as a polytrope:
\begin{equation}\label{eq:poly}
    P(s,\rho) = K(s)\rho^\Gamma = K(s)\rho^{1+1/n},
\end{equation}
where $P$ is the total pressure, $s$ is the entropy per baryon, $\rho$ is the mass density, and $\Gamma$ is the so-called polytropic exponent, with the associated quantity $n$ being the polytropic index. For a radiation dominated system, we can safely assume that the radiation field dominates the entropy per baryon $s\approx s_\text{rad.}=\tfrac{4}{3}m_\mathrm{b}aT^3/\rho$, where $T$ is the plasma temperature, and $m_\text{b}\approx931.5$ MeV is the mass of an atomic mass unit. Here, in natural units, $a=\pi^2/15$ is the radiation constant. Since $s$ is constant throughout the fully convective SMS, we can use Eq.~\eqref{eq:beta} and $P_\text{gas}=\rho T/(\mu m_\mathrm{b})
$ and $P_\text{rad.}=aT^4/3$ to derive
\begin{equation}
    s \approx \frac{4}{4.3}\left(\frac{M}{M_\odot}\right)^{1/2} - \frac{4}{\mu},
\end{equation}
where here we express $s$ in units of Boltzmann's constant $k_\text{B}$ per baryon. Note that for the SMSs considered here the entropy per baryon is large, e.g., $s\sim1000$ for $M\sim10^6\,M_\odot$. This large entropy implies a large number of photons per baryon, resulting in implications for the $e^\pm$ content of the star during its eventual collapse which, in turn, has implications for neutrino emissivity \cite{dicus,schinder,shifuller}. 

Using $P(s,\rho)=P_\text{gas}+P_\text{rad.}$ and $n=3$, $K(s)$ is found to be
\begin{align}
    K(s) &= \frac{1}{1-\beta}\frac{a}{3}\left(\frac{3}{4a}\frac{s}{\mu m_\text{b}}\right)^{4/3}, \nonumber \\ 
    &\approx\frac{0.5801}{1-\beta}\left(\frac{0.59}{\mu}\right)^{4/3}\left(\frac{s}{1000}\right)^{4/3}\,\rm{MeV}^{-4/3}.
\end{align}
For the stars discussed in this paper, their EOS and structure are well approximated by an $n=3$ ($\Gamma=4/3$) polytropic model. 

Since these stars are fully convective, we can take their entropy per baryon to be constant throughout. With constant entropy, the polytropic exponent $\Gamma=1+1/n$ is equivalent to Chandrasekhar's adiabatic index $\Gamma_1$, defined as~\cite{cg1}
\begin{equation}
    \Gamma_1 \equiv \left(\frac{\partial \log P}{\partial \log \rho}\right)\Big\vert_s.
    \label{gamma1}
\end{equation}

In the purely Newtonian gravitation case, the frequency of oscillations in response to radial perturbations is proportional to $(\left<\Gamma_1\right>_P - 4/3)^{1/2}$, where the pressure-averaged value of $\Gamma_1$ throughout the star is $\left<\Gamma_1\right>_P$. Therefore, in order for oscillations to have a real frequency and thereby stable against perturbations, $\left<\Gamma_1\right>_P$ must be strictly greater than $4/3$. When including \nth{1} Order GR corrections, however, this condition of stability changes to
\begin{equation}
    \left<\Gamma_1\right>_P > \frac{4}{3} + \bigO\left(\frac{R_\text{S}}{R}\right),
\end{equation}
where $R$ is the radius of the star and $R_\text{S} = 2GM$ is the star's Schwarzschild radius. For the types of stars considered in this paper, this correction is typically no more than one part in ten-thousand. 

As the star quasistatically contracts, however, $\left<\Gamma_1\right>_P$ asymptotically approaches $4/3$ from above, meaning that the adiabatic index will eventually fall below this threshold and make the whole configuration unstable. This point highlights a striking fact about the nature of these stars: while the internal structure is accurately determined by just Newtonian gravitation, their stability is entirely determined by GR. 

To quantify this instability point, we can extremize the total energy of the star. As a function of the central density $\rho_\text{c}$, the total energy $E$ of this star is~\cite{st}
\begin{equation}\label{eq:E}
    E(\rho_\text{c}) = k_1 M K \rho_\text{c}^{1/n} - k_2 G M^{5/3} \rho_\text{c}^{1/3} - k_4 G^2 M^{7/3}\rho_\text{c}^{2/3},
\end{equation}
where $k_1\approx1.7558$, $k_2\approx0.6390$, and $k_4\approx0.9183$ are structure coefficients calculated for an $n=3$ polytrope (see~\citet{st} for derivation). The first two terms of Eq.~\eqref{eq:E} are the standard internal gas energy and Newtonian gravitational binding energy, respectively, and the final term represents the non-linear gravitational self coupling term arising from \nth{1} Order GR corrections. A convenient dimensionless re-scaling follows from performing the transformations
\begin{subequations}\label{eq:trans}
    \begin{alignat}{3}
      \tilde E&= \frac{G^{3/2}}{K^{\Tilde{n}/2}}E, \label{eq:etilde}\\
      \tilde M&= \frac{G^{3/2}}{K^{\Tilde{n}/2}}M, \\
      x &= G M^{2/3}\rho_\text{c}^{1/3}, \label{eq:x}
    \end{alignat}
  \end{subequations}
where here $\Tilde{n}=1/(\Gamma_1-1)$ and we take the standard adiabatic index for a radiation pressure dominated plasma to be~\cite{g1}:
\begin{equation}
    \Gamma_1 \approx \frac{4}{3} + \frac{\beta}{6}.
\end{equation}
We assume $\Gamma_1$ is constant throughout the SMS since it is fully convective. Substituting the above into Eq.~\eqref{eq:E} gives
\begin{equation}\label{eq:Exb}
    \tilde E = k_1 \tilde M^{1/3-\beta/3}x^{1+\beta/2} - k_2 \tilde M x - k_4\tilde M x^2.
\end{equation}

The extremization follows on finding the first and second derivatives of Eq.~\eqref{eq:Exb} with respect to $x$, setting them both to zero, and solving simultaneously. The first derivative vanishing determines the point of hydrostatic equilibrium. The second derivative also vanishing determines where this hydrostatic configuration changes from stable to unstable. If we ignore the correction from GR and take the $\beta\rightarrow0$ limit for the first derivative, we find the standard Newtonian rescaled mass for an $n=3$ polytrope:
\begin{equation}\label{eq:mtilde}
    \tilde M = \left(\frac{k_1}{k_2}\right)^{3/2} \approx4.5547,
\end{equation}
which is independent of $x$ (and hence the central density) as expected for an $n=3$ polytrope. In other words, a star with Newtonian gravitation and supported entirely by radiation (particles with relativistic kinematics) has zero total energy, and hence there is no energy cost for expanding or contracting this configuration. This self gravitating configuration is therefore very close to instability.

Such metastability means that even very small effects can have a dramatic effect on stability to push a configuration over the edge. As we have argued above, GR gives a negligible effect on the structure of the star, i.e., the run of pressure with density. However, the inclusion of a very simple \nth{1}-Order correction from GR, namely one that captures the non-linearity of gravitation through self-coupling, can make stable configurations in Newtonian gravitation unstable.

Including general relativistic corrections in this way and simultaneously solving the two extremization equations, leads to a critical value of $x$ where a hydrostatic configuration changes from stable to unstable:
\begin{equation}
    x_\text{crit.} = \frac{k_2}{4k_4}\beta + \bigO(\beta^2).
\end{equation}
This corresponds to a critical central density of 
\begin{align}
    \rho_{\rm{c,\,crit.}}^0 &\approx \left(\frac{k_2}{4k_4}\right)^3\frac{\beta^3}{G^3M^2} \nonumber \\
        &\approx  3.98\left(\frac{0.59}{\mu}\right)^3\left(\frac{{10}^5\,M_\odot}{M}\right)^{7/2}\,{\rm g}\,{\rm cm}^{-3}\label{eq:rhoc}
\end{align}
where we have used Eq.~\eqref{eq:beta} to relate $\beta$ to the star's composition and mass. The physical interpretation of Eq.~\eqref{eq:rhoc} has two parts. First, note that the rest mass of an SMS stems almost entirely from baryons, whereas the pressure is mostly from radiation. Increasing the mean molecular weight $\mu$ by fusing baryons in composite nuclei has the effect of lowering the gas pressure contribution to the total pressure. That, in turn, decreases the critical density for the onset of instability. Secondly, increasing the mass $M$ of the SMS requires a higher pressure for hydrostatic equilibrium, implying a higher temperature and a higher pressure contribution from radiation. Hence, increasing mass also serves to push the critical central density lower.

\section{Collisionless Dark Matter}\label{sec:DM}

We will now consider the effects of including a cloud of collisionless DM that permeates the entire star. Assuming standard spherical accretion, we expect the density profile of this DM cloud to be given by~\cite{st}
\begin{equation}\label{eq:sphacc}
    \rho_{\text{DM}}(r) = \rho_{\text{DM}}^\infty\left(1 + \frac{2GM(r)}{v_\infty^2 r}\right)^{1/2}
\end{equation}
where $\rho_\infty$ and $v_\infty$ are the typical DM density and DM particle velocities far from any gravitational sources, respectively, and $M(r)$ is the total mass enclosed in a sphere of radius $r$. 


The final term in the above equation represents the ratio of the gravitational potential of a DM particle to its specific kinetic energy. It permits two natural limits: $\tfrac{1}{2}v_\infty^2\gg GM(r)/r$ and $\tfrac{1}{2}v_\infty^2\ll GM(r)/r$ which, following Ref.~\onlinecite{mnras}'s convention, we will call ``hot'' and ``cold'' DM, respectively. It should be noted that this nomenclature is not to be confused with the cosmological notion of ``Hot Dark Matter'' and ``Cold Dark Matter,'' which refers to the kinematics of DM at its decoupling epoch and how these kinematics affect early structure formation. Here, we will assume that both our ``hot'' and ``cold'' limits refer to DM particles with non-relativistic kinematics. 

\subsection{``Hot'' Dark Matter}\label{ss:HDM}

In the ``hot'' DM limit, Eq.~\eqref{eq:sphacc} reduces to the simple relation
\begin{equation}
    \rho_{\text{DM}}(r) = \rho_{\text{DM}}^\infty \equiv \rho_{\text{DM}},
\end{equation}
i.e.\ a constant density fluid to first order. The energy added to the system by the DM is calculated by first determining the gravitational potential created by the fluid. From Poisson's Equation, this DM has the harmonic potential
\begin{equation}
    \Phi_{\text{HDM}}(r) = \frac{2\pi}{3}G\rho_{\text{DM}}r^2 + C,
\end{equation}
where we have assumed regularity at the origin and C is a constant of integration. To calculate the total gravitational energy contributed by this fluid, we use
\begin{align}
    E_{\text{HDM}} &= \int\dd m \,\Phi_{\text{HDM}} \nonumber \\
           &= \frac{8\pi^2}{3}G\rho_{\text{DM}}\int_0^R\rho(r)r^4\,\dd r + C M.\label{eq:EHDMrho}
\end{align}
We will assume that $M$, $\dd m$, and $\rho(r)$ are dominated by the baryon rest mass. Note that $\rho(r)$ very closely follows the profile of an $n=3$ polytrope.  We will extremize the total energy while keeping the mass $M$ fixed, so the final term in Eq.~\eqref{eq:EHDMrho} will be ignored. The first term in Eq.~\eqref{eq:EHDMrho} can be evaluated by employing the well known solutions to the Lane-Emden Equation of order 3, yielding
\begin{equation}
    E_{\text{HDM}} = k_{\text{HDM}} M K \rho_{\text{DM}} \rho_\text{c}^{-2/3},
\end{equation}
where $k_{\text{HDM}} = 3.5845$ is the structure coefficient for the included DM and $K$ is the same from Eq.~\eqref{eq:poly}. Following the same transformation in  Eqs.~\eqref{eq:etilde}-\eqref{eq:x}, the total energy now reads
\begin{multline}
    \tilde E = k_1 \tilde M^{1/3-\beta/3}x^{1+\beta/2} - k_2 \tilde M x - k_4\tilde M x^2 \\+ k_{\text{HDM}}\tilde M^{1/3} x_{\text{DM}}^3x^{-2},
\end{multline}
where $x_\mathrm{DM}^3=G^3M^2\rho_\mathrm{DM}$. Extremization of this yields the following set of equations to be solved simultaneously after expanding in powers of the ratio of gas pressure to total pressure~$\beta$:
\begin{subequations}
\begin{alignat}{2}
    0 &= \frac{k_1}{\Tilde{M}^{2/3}}\left(1-2\frac{k_{\text{HDM}}}{k_1}\frac{x_{\text{DM}}^3}{x_\text{crit.}^3}\right) - k_2 - 2k_4x_\text{crit.} + \mathcal{O}(\beta), \\
    0 &= -2k_4 + \frac{k_1}{2\Tilde{M}^{2/3}}\frac{\beta}{x_\text{crit.}}\left(1+12\frac{k_{\text{HDM}}}{k_1\beta}\frac{x_{\text{DM}}^3}{x_\text{crit.}^3}\right) + \mathcal{O}(\beta^2).
\end{alignat}
\end{subequations}
After identifying that the ratio $x_{\text{DM}}^3/x^3$ is related to the DM mass fraction $f$ via
\begin{equation}\label{eq:xtof}
    \frac{x_{\text{DM}}^3}{x_\text{crit.}^3} = \frac{\rho_{\text{DM}}}{\rho_\text{c,\,crit.}} = \frac{\rho_{\text{DM}}}{C_3\rho_\text{avg.}} = \frac{M_{\text{DM}}}{C_3M} = \frac{f}{C_3},
\end{equation}
where $C_3\approx54.18$ is the core-to-average-density ratio (a.k.a\ ``condensation parameter'') for a $n=3$ polytrope, we find that
\begin{equation}
    \rho_\text{c,\,crit.} \approx \frac{1}{G^3M^2}\left(\frac{k_2}{4k_4}\beta + \frac{3k_1k_{\text{HDM}}}{k_2k_4}\frac{f}{C_3}\right)^3,
\end{equation}
which for typical values of $\mu$, $M$, and $f$ is
\begin{multline}\label{eq:rhocDM}
            \rho_\text{c,\,crit.}^\text{HDM} \approx \rho_\text{c,\,crit.}^0 \times \\ \left[1+0.197\left(\frac{\mu}{0.59}\right) \left(\frac{M}{10^5\,M_\odot}\right)^{1/2}\left(\frac{f}{0.01}\right)\right]^3,
\end{multline}
where $\rho_\text{c,\,crit.}^0$ is the critical central density without DM given in Eq.~\eqref{eq:rhoc}. Here, we have ignored $\bigO(\beta^4)$ and $\bigO(f^4)$ terms. It should be noted that all higher order terms in $f$ have positive coefficients, so Eq.~\eqref{eq:rhocDM} give a slight underestimate to the true increase in $\rho_\text{c, crit.}$ for larger $f$. Eq.~\eqref{eq:rhocDM} recovers the standard case when $f\rightarrow0$, as expected. 

\begin{figure}
    \centering
    \includegraphics[width=\linewidth]{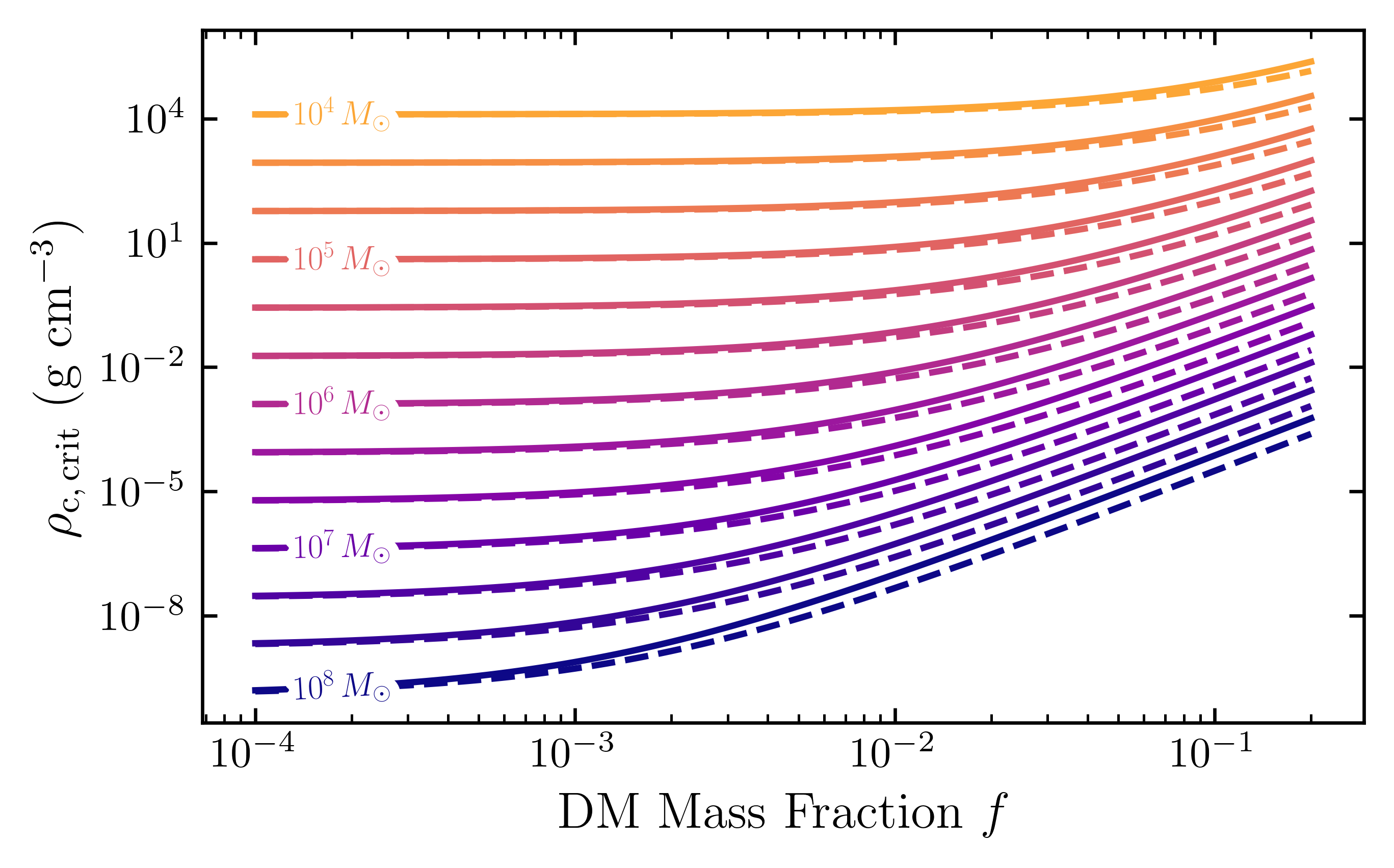}
    \caption{Critical central density for the onset of general relativistic instability, $\rho_{\rm c,crit}$, for SMSs vs.\ total DM mass fraction $f$ for different SMS masses. Dashed lines are the ``hot'' DM limit and solid lines are the ``cold'' DM limit (Eqs.~\eqref{eq:rhocDM} and \eqref{eq:rhocCDM}, respectively).}
    \label{fig:sms}
\end{figure}

In Figure \ref{fig:sms}, we show the critical central density $\rho_\text{c,\,crit.}$ at the onset of GR instability as a function of DM mass fraction $f$ in the hot DM case (dashed lines) for various SMS masses between $10^4$ and $10^8\,M_\odot$. While the relative increase is not too significant for smaller mass SMSs, the effect is quite strong for larger masses (e.g.~${\gtrsim}~10^5\,M_\odot$), increasing $\rho_\text{c,\,crit.}$ by as much as 4 orders of magnitude at $f\sim20\%$.

\subsection{``Cold'' Dark Matter}\label{ss:CDM}

The alternative limit of Eq.~\eqref{eq:sphacc} reduces to the profile
\begin{equation}
    \rho_{\text{DM}}(r) = \frac{\rho_\text{DM}^\infty}{v_\infty}\sqrt{\frac{GM}{r}},
\end{equation}
where we have also made the same approximation as \citet{mnras} that since an $n=3$ polytrope is so centrally condensed, it is reasonable to treat the SMS as a point source of mass $M$. Without this assumption, the DM density profile would feature a flattened core as opposed to the above divergent cusp at the origin. Such a flat shape would be closer to the limit described in Section~\ref{ss:HDM}, so the `cuspy' profile offers sufficiently disparate extreme from the ``hot'' DM case to allow us to explore the full range of DM effects we would expect to see. 

The mass fraction of this DM is found by integrating the density throughout the entire star and dividing by the mass:
\begin{align}
    f = \frac{M_\text{DM}}{M} &= 4\pi\sqrt{\frac{2G}{M}}\frac{\rho_\text{DM}}{v_\infty}\int_0^R\dd r \,r^{3/2}, \nonumber \\
    &= \frac{8\pi\sqrt{2}}{5}\sqrt{\frac{GR^5}{M}}\frac{\rho_\text{DM}}{v_\infty}, \nonumber \\
   f &= C_\text{CDM}\frac{x_\text{DM}^3}{x^{5/2}},\label{eq:fcdm}
\end{align}
where $C_\text{CDM}\approx60.0151$ is the ``cold'' DM equivalent to the condensation parameter in the previous section and is calculated in part from solutions to the Lane-Emden equation of order 3, $x$ is the reparameterized central baryonic density given by Eq.~\eqref{eq:x}, and $x_\text{DM}$ is defined as
\begin{equation}\label{eq:xCDM}
    x_\text{DM}^3 = G^3M^2\frac{\rho_\text{DM}^\infty}{v_\infty}.
\end{equation}
This definition is similar to $x_\text{DM}$ of the previous section with the inclusion of $v_\infty$, the typical DM particle velocity far from the SMS. Since the only the ratio $\rho_\text{DM}^\infty/v_\infty$ appears in our calculations, both have been collapsed into the single quantity $x_\text{DM}$.

Solving Poisson's Equation again for the gravitational potential caused by this ``cold'' DM, we find that
\begin{equation}
    \Phi_\text{CDM}(r) = \frac{16\pi\sqrt{2}}{15}G^{3/2}M^{1/2}\frac{\rho_\text{DM}^\infty}{v_\infty}r^{3/2},
\end{equation}
where we again assume regularity at the origin and have ignored the constant of integration for the same reason as the previous section. The total gravitational energy is then calculated to be
\begin{align}
    E_\text{CDM} &= \int\dd m\,\Phi_\text{CDM} \nonumber \\
    &= k_\text{CDM} M K^{3/2}\frac{\rho_\text{DM}^\infty}{v_\infty}\rho_\text{c}^{-1/2}
\end{align}
where $k_\text{CDM}\approx14.066$ is calculated following a similar process to the previous section. The rescaled energy per Eqs.~\eqref{eq:etilde}-\eqref{eq:x} and Eq.~\eqref{eq:xCDM} is then
\begin{equation}
    \tilde E_\text{CDM} = k_\text{CDM}x_\text{DM}^3x^{-3/2},
\end{equation}
so the total rescaled energy is
\begin{multline}
    \tilde E = k_1 \tilde M^{1/3-\beta/3}x^{1+\beta/2} - k_2 \tilde M x - k_4\tilde M x^2 \\+ k_\text{CDM}x_\text{DM}^3x^{-3/2}.
\end{multline}

Following the same process as the ``hot'' DM case with the mass fraction in Eq.~\eqref{eq:fcdm}, we find that
\begin{equation}
    \rho_\text{c,\,crit.}\approx\frac{1}{G^3M^2}\left(\frac{k_2}{4k_4}\beta + \frac{15k_2k_\text{CDM}}{8 \tilde M^{1/3} k_1 k_4}\frac{f}{C_\text{CDM}}\right)^3,
\end{equation}
where $\tilde M=4.5547$ (per Eq.~\eqref{eq:mtilde}) and we have again ignored \nth{4} order terms in $\beta$ and $f$. Again, for typical values of $\mu$, $M$, and $f$, this reduces to
\begin{multline}\label{eq:rhocCDM}
    \rho_\text{c,\,crit.}^\text{CDM} \approx \rho_\text{c,\,crit.}^0 \times \\ \left[1+0.262\left(\frac{\mu}{0.59}\right) \left(\frac{M}{10^5\,M_\odot}\right)^{1/2}\left(\frac{f}{0.01}\right)\right]^3,
\end{multline}
which has a slightly stronger dependence on the DM mass fraction than the ``hot'' DM case, as evidenced in Figure \ref{fig:sms}.

\section{DM Response to SMS Contraction}\label{sec:response}

As the star quasistatically contracts, we would like to quantify how much the mass fraction of DM within the star changes. If the fraction of DM decreases for a hypothetical decrease in radius (enclosing the baryons), the SMS would be enclosing less total mass. That may help stabilize the configuration. Alternatively, or perhaps in addition to this effect, if SMS contraction serves to increase the velocity dispersion of the DM, the effective pressure provided by the dispersion would increase and also stabilize the configuration.

From Eqs.~\eqref{eq:xtof} and \eqref{eq:fcdm}, the fractional change in DM fraction $f$ is
\begin{equation}
    \frac{\delta f}{f} = \frac{\delta\rho_{\text{DM}}}{\rho_{\text{DM}}} - \kappa\frac{\delta\rho_\text{c}}{\rho_\text{c}},
\end{equation}
where $\kappa=1$ or $5/6$ for the ``hot'' and ``cold'' limits, respectively.

The largest potential deviation in DM density from its background value would occur where $M(r)/r$ achieves its maximum value, which is at $r=R$, so we will consider that region and small changes in the SMS's radius $\delta R < 0$ where $\delta R/R\ll1$. Since $\rho_\text{c}\sim M/R^3$,
\begin{equation}
    \frac{\delta\rho_\text{c}}{\rho_\text{c}} = -3\frac{\delta R}{R}
\end{equation}
for a fixed mass, and from Eq.~\eqref{eq:sphacc} we find that
\begin{equation}\label{eq:fracdeltaDM}
    \frac{\delta\rho_{\text{DM}}}{\rho_{\text{DM}}} = -\frac{GM}{v_\infty^2R}\left(1 + \frac{2GM}{v_\infty^2R}\right)^{-1}\frac{\delta R}{R}.
\end{equation}
We will now consider both the ``hot'' and ``cold'' DM limits to calculate this fractional change. 

\subsection{``Hot'' Dark Matter Case}

To first order in $GM/(v_\infty^2R)$, Eq.~\eqref{eq:fracdeltaDM} reduces to
\begin{equation}
    \frac{\delta \rho_{\text{DM}}}{\rho_{\text{DM}}} \approx - \frac{GM}{v_\infty^2R}\frac{\delta R}{R},
\end{equation}
so
\begin{equation}\label{eq:fracdfHot}
    \frac{\delta f}{f} \approx 3\left(1 - \frac{GM}{3v_\infty^2R}\right)\frac{\delta R}{R}.
\end{equation}
Therefore, the total mass fraction of DM would tend to decrease since $\delta R < 0$, unless $GM/(6R) > \tfrac{1}{2}v_\infty^2$ which is the opposite of the limit we are considering, so this can be further reduced to
\begin{equation}
    \left(\frac{\delta f}{f}\right)_\text{HDM} \approx 3\frac{\delta R}{R}.
\end{equation}

\subsection{``Cold'' Dark Matter Case}

If $GM/(v_\infty^2R)\gg1$, then Eq.~\eqref{eq:fracdeltaDM} becomes simply
\begin{equation}
    \frac{\delta\rho_{DM}}{\rho_{DM}} \approx -\frac{1}{2}\frac{\delta R}{R},
\end{equation}
so
\begin{equation}\label{eq:fracdfCold}
    \left(\frac{\delta f}{f}\right)_\text{CDM} \approx 2\frac{\delta R}{R},
\end{equation}
which is again negative for $\delta R<0$, indicating that the total DM mass fraction also decreases in this limit but by a less severe amount compared to the ``hot'' DM case.

\subsection{Physical Origin of Enhanced Stability}

The stability of a purely standard model star (i.e., composed of baryons, electrons, photons, etc.) is determined by its pressure-averaged adiabatic index as given in Eq.~\eqref{gamma1}. A heuristic physical picture for this effective \lq\lq spring constant\rq\rq\ for the star's material is as follows: Squeeze the star radially and rapidly (so rapidly that heat does not flow and so entropy remains constant) to higher density and ask how much the pressure increases \--- i.e., how much does the material \lq\lq push back\rq\rq\ when quickly perturbed to higher density? 

Of course, when the pressure support for a self-gravitating configuration stems from particles with relativistic kinematics (e.g., photons in the SMS case), that configuration is close to instability. For example, this case would correspond to a zero total energy configuration in purely Newtonian gravitation, therefore pressure forces will grow in lock-step with the increased gravitational force from compressing the star. This means the configuration is in a meta-stable state as it costs no energy to expand or contract the star, making it ripe for instability once the non-linearity of GR comes into play.

When collisionless DM is added to this picture there are several effects to consider. We envision that the standard model portion of the star, contained inside a radius $R$, is embedded in an extended halo of DM. When the star is radially and adiabatically squeezed to higher baryon density there are two questions that arise: (1) What is the effect of reducing $R$ in the limit where the DM distribution in space is fixed so that less DM is enclosed inside the SMS radius?; and (2) If the standard model star is squeezed to higher density, what is the response of the DM spatial distribution and how does this affect stability?  

To the first issue, as discussed previously, enclosing less total mass as the SMS is squeezed is a contributing factor in the enhanced stability afforded by DM. Indeed, as evidenced in Eqs.~\eqref{eq:fracdfHot} and \eqref{eq:fracdfCold}, the total DM mass fraction decreases in both DM kinematic limits. Curiously, this total mass decrement effect is less pronounced for the ``cold'' case despite it producing a stronger stabilizing effect (c.f.\ the coefficients in the correction terms in Eqs.~\eqref{eq:rhocCDM} and \eqref{eq:rhocDM} and Figure \ref{fig:sms}). This indicates that DM induced stabilization is not only due to enclosing less \emph{total} mass. 

As for the second issue, the gravitational response of the DM spatial distribution to a perturbation in the standard model star is second order, so it is not accounted for in our calculations. The ``cold'' DM case possesses a more centrally condensed DM distribution, which changes the outcome of the extremization calculation as compared to the ``hot'' case. In the end, adding non-relativistic stress from DM to the supermassive star in general, and doing so in a way that provides a nearly fixed background, is inherently stabilizing.

We note that the increased stability stemming from a fixed DM distribution inside our SMS may have similarity with the enhanced stability found in giant-planet envelope models with a fixed, incompressible core \cite{Kundu2021}. In both the giant-planet envelope and our SMS cases there is little or no change in the background gravitational potential of the core in the former case, or the DM envelope in the latter.

\section{Equilibrium Structure of Star with Dark Matter}\label{sec:laneemdenf}

Let us consider the ``hot'' dark matter case where the density profile follows a flat distribution. Considering that SMSs are well approximated by the structure of an $n=3$ polytrope, they are highly centrally condensed relative to their average density. For any non-zero DM mass fraction $f$, there should be a radius $r_\mathrm{eq.} < R$ beyond which the DM density is greater than the standard model matter density. Since the pressure gradient of the SMS depends on the local gravitational field strength, we should expect that around and beyond this radius the SMS structure may be altered by this extra gravitation source.

Specifically, if we assume the DM provides no direct pressure support but does interact with the SMS gravitationally, the pressure gradient at a given radius $r$ is
\begin{equation}\label{eq:dpdrq}
    \frac{\dd P_0}{\dd r} = -\frac{G(m_0 + m_\mathrm{DM})(\rho_0 + \rho_\mathrm{DM})}{r^2},
\end{equation}
where $P_0$, $m_0$, and $\rho_0$ are the standard model pressure, mass enclosed, and density, respectively, and $m_\mathrm{DM}$ and $\rho_\mathrm{DM}$ are the DM mass enclosed and density, respectively. Both $m_0$ and $m_\mathrm{DM}$ have the following continuity conditions:
\begin{align}
    \frac{\dd m_0}{\dd r} &= 4\pi r^2\rho_0, \\
    \frac{\dd m_\mathrm{DM}}{\dd r} &= 4\pi r^2\rho_\mathrm{DM}.
\end{align}

With these three equations, Eq.~\eqref{eq:dpdrq} can be manipulated to derive
\begin{equation}\label{eq:poisson}
    \frac{1}{r^2}\frac{\dd}{\dd r}\left(\frac{r^2}{\rho_0+\rho_\mathrm{DM}}\frac{\dd P_0}{\dd r}\right) = -4\pi G(\rho_0+\rho_\mathrm{DM}).
\end{equation}
Let us now assume that the standard model density is parameterized by its central density $\rho_{\rm 0c}$ and a function of radius $\theta$ via $\rho_0 = \rho_{\rm 0c}\theta^n$, where $n$ is the polytropic index. Plugging this into Eq.~\eqref{eq:poly} yields $P_0 = K\rho_{\rm 0c}^{1+1/n}\theta^{n+1}$. Collecting these relations together in Eq.~\eqref{eq:poisson} gives
\begin{equation}
    \frac{1}{r^2}\frac{\dd}{\dd r}\left(\frac{r^2}{\theta^n+\frac{f}{C_3}}\alpha_\mathrm{LE}^2\theta^n\frac{\dd \theta}{\dd r}\right) = -\left(\theta^n + \frac{f}{C_3}\right),
\end{equation}
where $\alpha_\mathrm{LE}^2 = (n+1)K\rho_{\rm 0c}^{1/n-1}/(4\pi G)$ is the square of the standard Lane-Emden length scale and $f/C_3$ is the DM mass fraction scaled by the central condensation as per Eq.~\eqref{eq:xtof}, specifically for a flat DM distribution.

\begin{figure}[h]
    \centering
    \includegraphics[width=\linewidth]{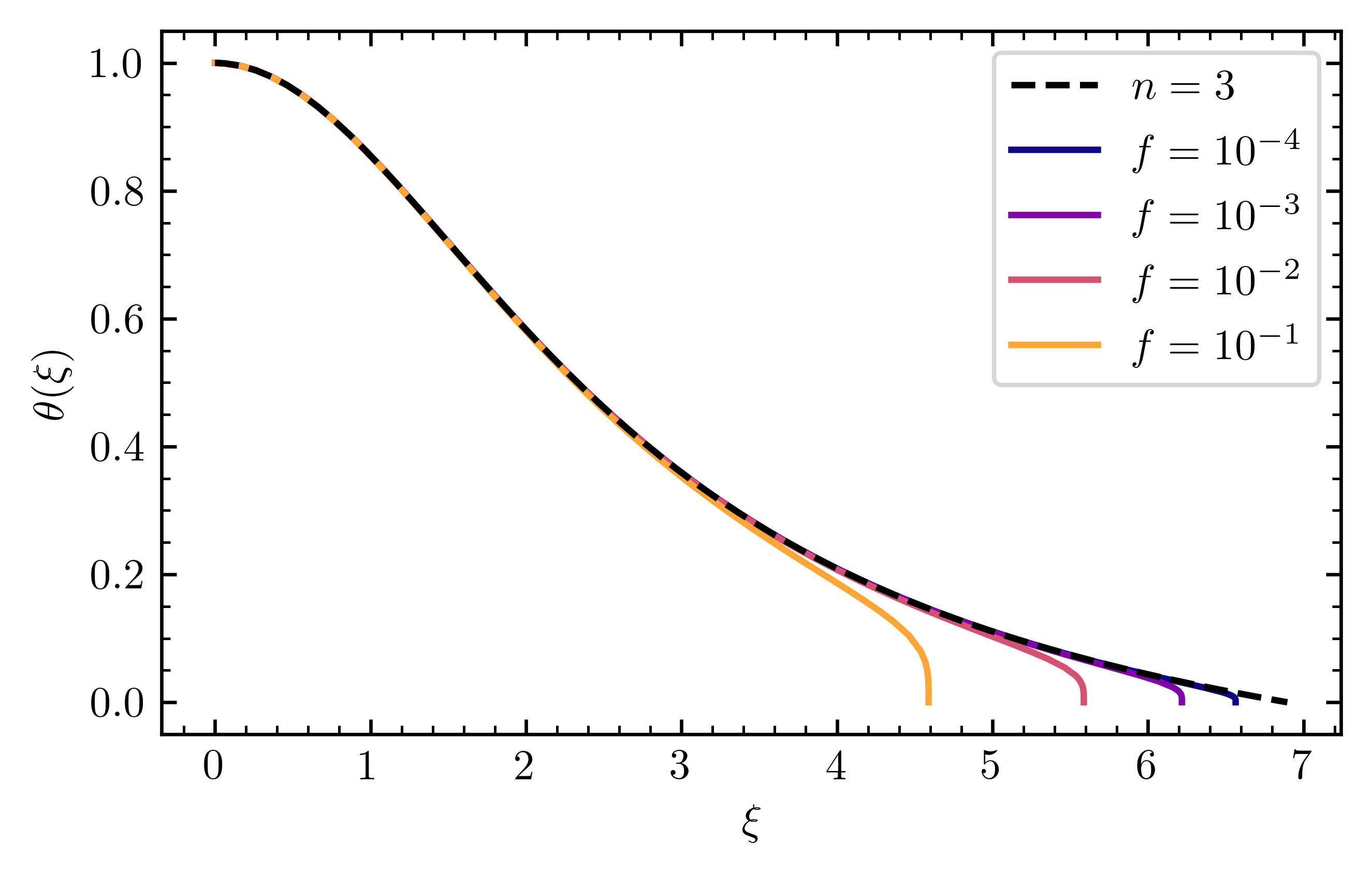}
    \caption{Solutions to Eq.~\eqref{eq:laneemdenf} of the function $\theta$ for different DM mass fractions $f$. The actual physical density is $\rho_0(r)=\rho_{\rm c0}\theta^3(\xi)$, where $\xi=r/\alpha_\mathrm{LE}$.}
    \label{fig:laneemdenf}
\end{figure}

Scaling the radius $r$ via $r=\alpha_\mathrm{LE}\xi$, we arrive at a modified Lane-Emden equation which reads
\begin{equation}\label{eq:laneemdenf}
    \frac{1}{\xi^2}\frac{\dd}{\dd\xi}\left(\frac{\xi^2}{\theta^n+\frac{f}{C_3}}\theta^n\frac{\dd\theta}{\dd\xi}\right) = -\left(\theta^n + \frac{f}{C_3}\right),
\end{equation}
which reduces to the standard Lane-Emden equation for $f\rightarrow0$. We can solve this equation numerically for different values of $f$, as presented in Figure~\ref{fig:laneemdenf} for $n=3$. The solutions agree with the standard (no DM) case for most of the x-axis range of the figure, but begin to diverge radically at roughly where the standard model and DM densities are comparable. 

\begin{figure}[h]
    \centering
    \includegraphics[width=\linewidth]{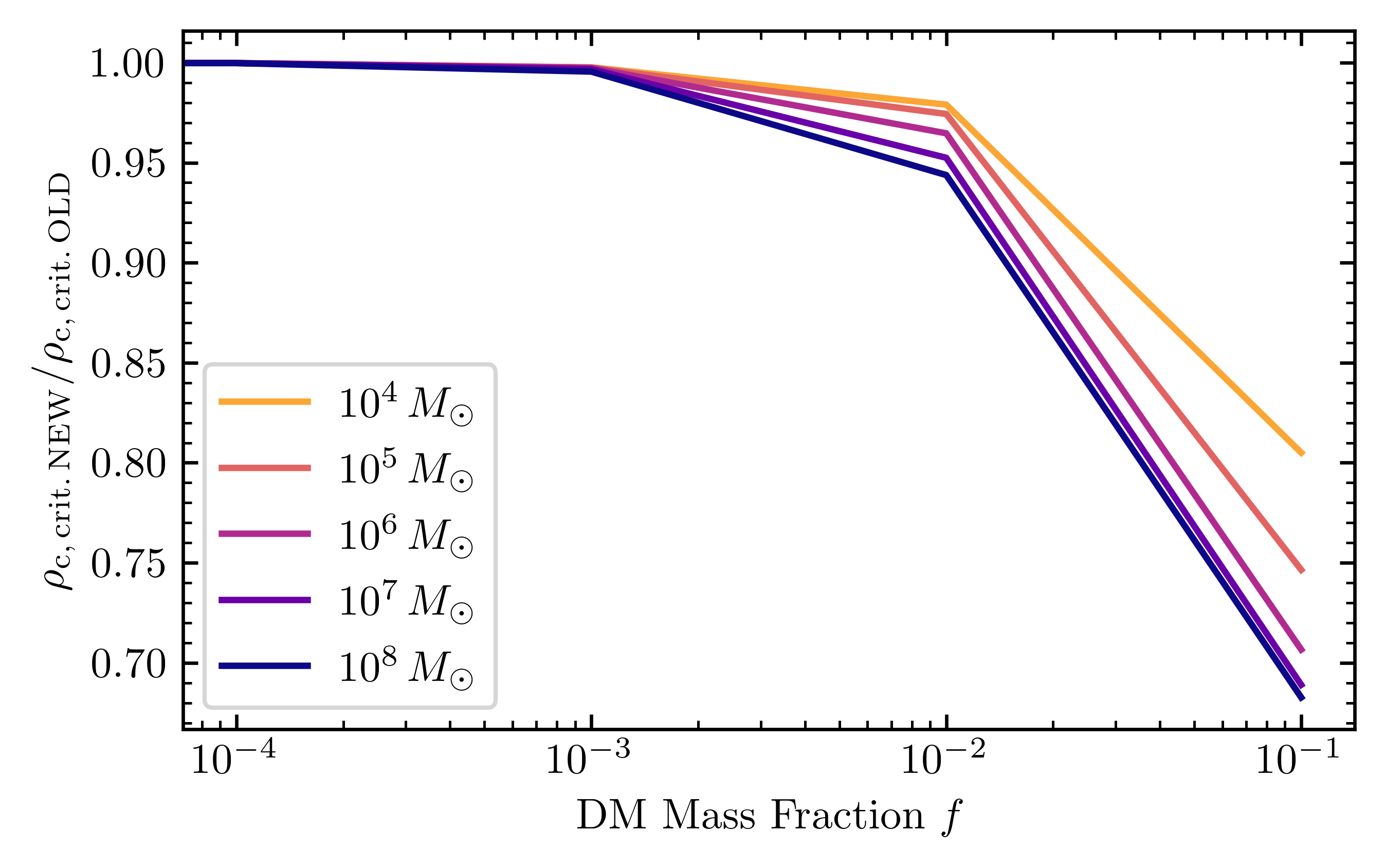}
    \caption{As a function of the DM mass fraction $f$, the relative change in the critical central density is plotted. Here, the quantity $\rho_\mathrm{c,\,crit.NEW}/\rho_\mathrm{c,\,crit.OLD}$ is the ratio of the critical central densities considering the change in the equilibrium density profile of the SMS \emph{vs.}~not considering such a change.}
    \label{fig:rhoccritf}
\end{figure}

Using these functions, the altered values of the structure coefficients $k_1$, $k_2$, $k_4$, and $k_\mathrm{HDM}$ are tabulated as a function of DM mass fraction $f$ in Table~\ref{tab:coeffs}. These were calculated following the prescriptions in Chapter 6 of \citet{st}. In Figure~\ref{fig:rhoccritf}, the relative change in the critical central density for several SMS masses and DM mass fractions is visualized. For all SMS masses and DM mass fractions, including the effect of the altered equilibrium distribution of the standard model matter lowers the critical central density slightly from Eq.~\eqref{eq:rhocDM}. 

\begin{table}[h]
\begin{tabular}{@{}lccccc@{}}
\toprule
                 & $n=3$   & $f=10^{-4}$ & $f=10^{-3}$ & $f=10^{-2}$ & $f=10^{-1}$ \\ \midrule
$k_1$            & 1.75579 & 1.75595     & 1.75714     & 1.76558     & 1.80913     \\
$k_2$            & 0.63900 & 0.63911     & 0.63999     & 0.64650     & 0.68533     \\
$k_4$            & 0.91829 & 0.91853     & 0.92044     & 0.93485     & 1.02386     \\
$k_\mathrm{HDM}$ & 3.58450 & 3.58409     & 3.58044     & 3.55002     & 3.37575     \\ \bottomrule
\end{tabular}
\caption{Tabulated structure coefficients determined upon integrating the solutions to Eq.~\eqref{eq:laneemdenf}. The ``$n=3$'' column corresponds to the standard Lane-Emden solution (i.e. $f=0$), and the following columns are the values calculated for several chosen and increasing DM mass fractions $f$.}
\label{tab:coeffs}
\end{table}

Clearly, considering the change in the SMS equilibrium density profile only confers significant changes in the critical central density for DM mass fractions $f$ in the larger end of the range we are considering (${\gtrsim}1\%$).

\section{Nuclear Burning and Neutrino Losses}\label{sec:ppi}

It should be expected that, at least for some masses in the range ${\sim}10^4\text{--}10^8\,M_\odot$, nuclear (hydrogen) burning on the PPI chain is taking place at a high enough rate to significantly contribute to the total energy budget of the SMS. This rate depends on both the density and temperature throughout the star, both of which decrease as the mass increases. Specifically, \emph{at the onset of collapse} where the density and temperature are highest, $\rho_\text{c}\sim M^{-7/2}$ and $T_c\sim M^{-1}$, the latter following from the definition of an $n=3$ polytrope. Since including DM can raise the critical central density (and hence the temperature) required before collapse, we can also investigate how the DM mass fraction impacts the energy released by the PPI chain throughout the star. \citet{Chen_2014} investigates Helium burning in SMSs, which could be relevant for SMSs light enough to have a stable Hydrogen burning phase. The following calculations in this section are determined for the non-altered SMS density distribution, \emph{contra} the previous section's treatment, for simplicity.

While significant nuclear burning would potentially lead the star through a stable ``main-sequence'' phase of evolution, neutrino emission from the high entropy plasma could carry that extra energy generation away from the core, providing no direct pressure support and hence adding no more stability to the configuration. At such high entropies and temperatures, the neutrino luminosity may outshine that of nuclear burning completely, causing the star to contract more quickly and leading it closer to GR instability. 

\subsection{PPI Chain Nuclear Burning}

For zero metallicity, protons $p$ can only fuse to Helium nuclei $\alpha$ via the PPI chain $4p\rightarrow\alpha+2e^+ + 2\nu_e$, where $e^+$ indicates positrons and $\nu_e$ indicates electron neutrinos. In total, each such reaction releases about $26$ MeV of energy to the plasma while the neutrinos carry away about $0.5$ MeV away from the star as the likelihood they will transfer energy to the plasma is vanishingly small~\cite{clayton}. The energy generation rate for this reaction is calculated to be (in $\mathrm{erg\,cm^{-3}\,s^{-1}}$)
\begin{multline}
    Q_{\text{PPI}} = c_1\rho^2X_\text{H}^2T_6^{-2/3}\exp\left(-c_2 T_6^{-1/3}\right)\\
    \times\left(1+c_3T_6^{1/3} + c_4T_6^{2/3} + c_5T_6\right),
\end{multline}
where $\rho$ is the density in $\mathrm{g\,cm^{-3}}$, $X_\text{H}\approx0.75$ is the total hydrogen mass fraction, $T_6 = T/(10^6\,\mathrm{K})$ is the temperature in units of $10^6$ Kelvins, and ${c_1=2.319\times10^6}$, $c_2=33.81$, $c_3=0.0123$, $c_4=0.0109$, $c_5=0.00095$ are numerical constants (see \citet{clayton} for derivation). 

To determine if the PPI chain provides a significant contribution to the energy budget of the star, we will compare the integrated luminosity at the surface, i.e.
\begin{equation}
    L_\text{PPI} = \int_0^RQ_\text{PPI}4\pi r^2\dd r
\end{equation}
to the Eddington luminosity, which is approximately~\cite{st}
\begin{equation}
    L_\text{Edd.} \approx 1.3\times10^{38}\left(\frac{M}{M_\odot}\right)\,\mathrm{erg\,s^{-1}},
\end{equation}
which comes from considering at what photon luminosity momentum transfer via photon-electron Thomson scattering balances the gravitational force.

\subsection{Neutrino Losses}

The high entropy per baryon ($s\gtrsim1000$) required for hydrostatic equilibrium in SMSs implies a significant $e^\pm$ pair content in electromagnetic equilibrium $e^+~+~e^-~\leftrightarrow~2\gamma$. These electrons and positrons can both annihilate and interact with the local radiation field to produce neutrinos via~\cite{dicus,schinder,1989ApJ...339..354I}:
\begin{align*}
    e^- + e^+ &\rightarrow \nu + \Bar{\nu}\,&\text{(pair production)} \\
    \gamma + e^{\pm} &\rightarrow e^{\pm} + \nu + \Bar{\nu}\,&\text{(photo-emission)}
\end{align*}
where $\gamma$ represents an incident photon and $\nu$ ($\Bar{\nu}$) refers to (anti-)neutrinos of any flavor. These processes are expected to be the dominant source of neutrino emissivity in the low density and low-to-moderate temperature regimes in SMSs. There is also a process where plasmons ($\Tilde{\gamma}$) decay to neutrino pairs ($\Tilde{\gamma}\rightarrow\nu+\Bar{\nu}$), but this only dominates in high density and temperature regimes. 

If the electrons and positrons are non-relativistic and non-degenerate, the energy density loss rates for the two dominant neutrino production processes are calculated to be \cite{clayton,dicus} (in $\mathrm{erg\,cm^{-3}\,s^{-1}}$):
\begin{align}
    Q_\text{pair} &\approx \left(4.9\times10^{18}\right)T_9^3\exp\left(-\frac{11.86}{T_9}\right), \\
    Q_\text{photo.} &\approx \left(0.98\times10^8\right)\frac{\rho}{\mu}\,T_9^8,
\end{align}
where $\rho$ is the mass density in $\mathrm{g\,cm^{-3}}$, $\mu$ is the mean molecular weight, and $T_9=T/(10^9\,\mathrm{K)}$ is the temperature in units of $10^9$ Kelvins. While the pair production mechanism has a much higher prefactor than the photo-emission process, the former suffers an exponential suppression for low temperatures that the latter channel does not, actually making that channel dominant in higher mass SMSs.

Using a similar process as in the previous section, we calculate the combined luminosity from these neutrino emission processes to be
\begin{equation}
    L_\mathrm{\nu,\,Tot.} = \int_0^R\left(Q_\mathrm{pair}+Q_\mathrm{photo.}\right)4\pi r^2\,\dd r.
\end{equation}
While the PPI chain does emit two $\nu_e$ per reaction, the energy they carry away is far smaller than the energy released from fusing four protons into one $\alpha$ particle, so we will ignore that contribution from the total neutrino luminosity.

\begin{figure}
    \centering
    \includegraphics[width=\linewidth]{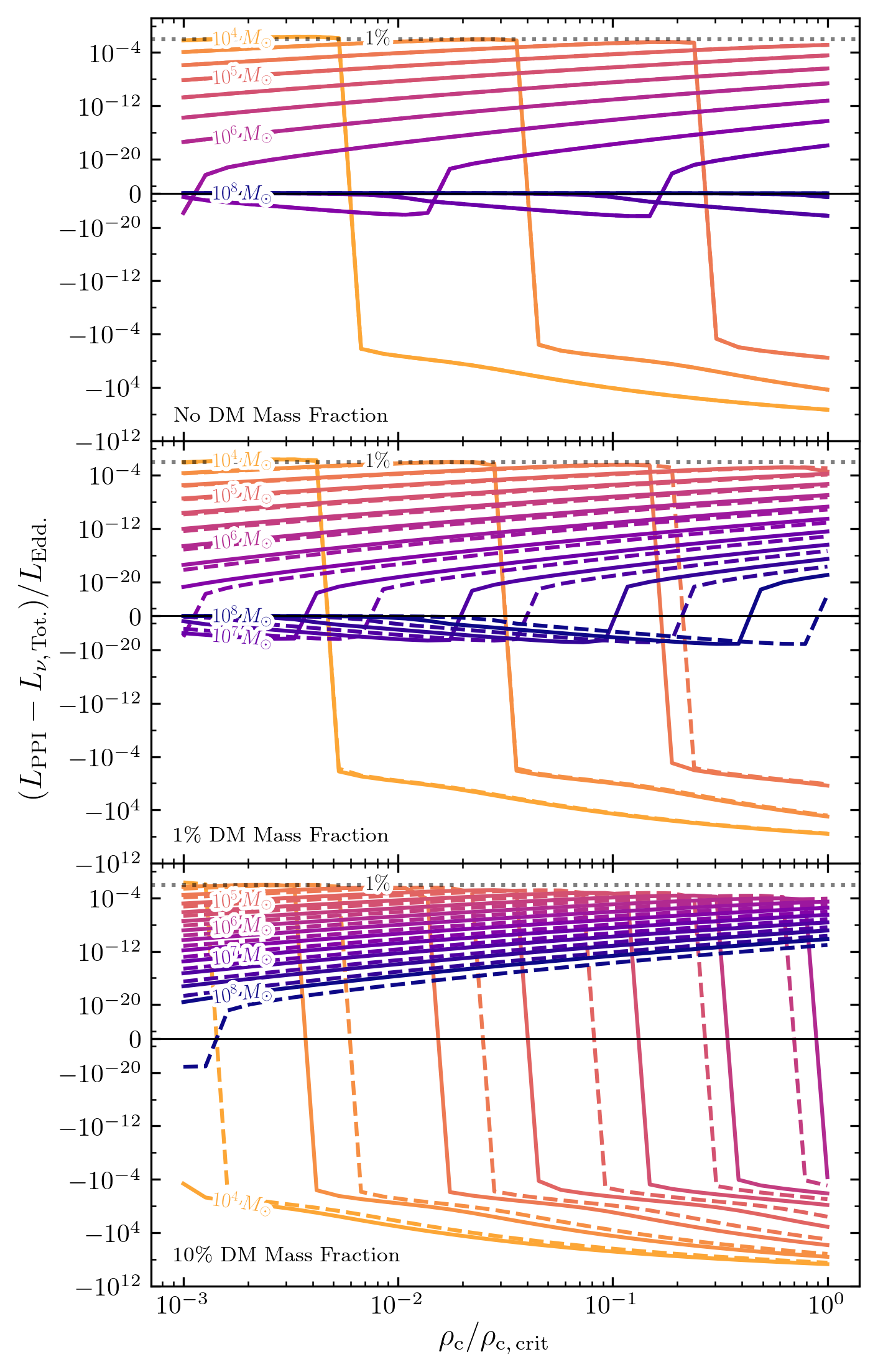}
    \caption{Symmetric log-log triptych plot showing total luminosity (burning minus neutrino losses) normalized to each SMSs' Eddington luminosity versus the SMS's central density as a fraction of its critical central density for different DM mass fractions. Solid (dashed) curves correspond to the ``cold'' (``hot'') kinematic limit. The top panel is for a star with no DM, the middle for a star with 1\% DM by mass, and the bottom for a star with 10\% DM by mass, all for various SMS masses between $10^4\text{--}10^8\,M_\odot$. Indicated in each panel is a dashed gray line at 1\% Eddington luminosity to indicate when burning is ``important'' for the evolution of the star. Note that these should not be taken to be evolutionary plots since the mean molecular weight $\mu$ is fixed.}
    \label{fig:adiabaticLong}
\end{figure}

To examine how the PPI chain luminosity compares to the neutrino luminosity as the star contracts, we can simply scale down the central density and let it approach the critical central density from below. Assuming that the run of density and temperature closely follows an $n=3$ polytrope, this is a valid analysis as such a polytrope is entirely self similar (i.e.\ the mass is independent of the central density and only depends on $s$ and $\mu$ via the polytropic constant of proportionality $K$). 

In Figure \ref{fig:adiabaticLong}, we present this result in three symmetric log-log plots showing the total luminosity from both sources ($L_\text{PPI}~-~L_\mathrm{\nu,\,Tot.}$) as a fraction of the Eddington luminosity for various SMS masses and DM mass fractions. We subtract the neutrino luminosity since it carries away energy from the star, as opposed to creating energy like the PPI process. This plot should not be treated as an evolutionary plot since the curves shown in the figure are calculated with a fixed mean molecular weight, $\mu=0.59$, i.e., the value corresponding to a fully ionized primordial gas. Nuclear evolution, in the sense of burning via incorporating baryons into nuclei, should serve to increase the mean molecular weight $\mu$ as higher mass nuclei are created. 

We would also expect that when nuclear burning is ``important'' (for the purposes of this paper chosen to be ${\sim}1\%$ of a star's Eddington luminosity) the central density is held almost constant until its fuel is exhausted. This would prevent an unabated contraction of the star to the critical central density for GR instability, if only for a relatively short duration.

Since adding DM serves to increase the critical central density, more DM can push massive stars to higher central densities without causing them to become unstable. Both the PPI process and the relevant neutrino processes depend sensitively on the central temperature (and hence the central density), so a delicate balancing act ensues between energy generation from proton fusion and energy loss from neutrino emission.

As evidenced from the top plot (i.e.\ zero DM) in Figure \ref{fig:adiabaticLong}, specifically for SMSs of masses ${\lesssim}10^5\,M_\odot$, increasing the central density increases the fraction of PPI energy generation, and hence net energy deposition, but only up to a point. Eventually, the neutrino losses from the increasing temperature win out, and the star is net losing energy instead of generating energy. 

Should the quasi-statically contracting SMS encounter this situation during its evolution, the neutrino emission would cause the star to contract more quickly. This is because the neutrinos carry away the plasma's entropy faster than Hydrogen burning can generate it throughout the configuration. Per the solutions to the Lane-Emden equations, an $n=3$ polytrope's radius scales as $R\sim K^{1/2}\sim s^{2/3}$, and since $\rho_\text{c}\sim R^{-3}$, the central density will increase as the radius decreases as neutrinos remove entropy from the SMS. This serves to push the SMS closer to instability.

The situation changes slightly for the ${\sim}1\%$ DM mass fraction case (middle panel) and quite dramatically for the ${\sim}10\%$ case (bottom panel). At a given central density fraction $\rho_\text{c}/\rho_\text{c,\,crit}$, increasing DM mass fraction corresponds to a higher absolute density on each mass curve. If we examine the $10^5\,M_\odot$ curve, for example, this mechanism allows for the energy generation fraction from PPI to quite closely approach $1\%$ of the Eddington luminosity as the DM mass fraction increases. This potentially permits such stars to burn stably on the main sequence instead of a direct collapse as expected in the standard case. This would drastically change their evolution as opposed to the case without DM. 

\section{Conclusion}

These SMSs are quite interesting case studies, primarily because their existence provides a potential physical explanation for several observational mysteries that have only deepened as more advanced telescopes have come online. Not only has JWST, as an example, seen plenty of high mass quasars at high redshift which puts the standard picture of solar mass remnant accretion into doubt, it may also have observed SMSs supported in some way by DM prior to collapse in the JADES survey, per \citet{ds}. 

Considering that these stars are ``trembling on the verge of instability'' due to the Feynman-Chandrasekhar Instability, it begs the question as to how they would be observed at all prior to collapse. One possible avenue to support these stars is through some form of non-interacting and non-relativistic DM. The enhanced stability would certainly also have implications for the nuclear and neutrino evolution of the stars.

In Section \ref{sec:stability}, we derived the standard result for the critical central density beyond which a hydrostatic SMS is unstable to collapse. This rested upon extremizing the total energy of the star with respect to the central density. We found that, as expected, higher mean molecular weight plasmas and higher mass SMSs are more unstable to collapse. Both of these factors serve to decrease the critical central density.

In Sections \ref{ss:HDM} and \ref{ss:CDM}, we derived analytically how the critical central density changes as a function of the DM mass fraction within the star. We considered two limiting cases of the DM kinematics, the ``hot'' case and the ``cold'' case, which compared the typical DM particle's specific kinetic energy far away from the SMS to the gravitational potential felt at the SMS surface, and found that in both cases the critical central density for the onset of general relativistic instability increased as the DM mass fraction increases. The ``cold'' case shows a slightly stronger effect, and the relative increase is greater for more massive stars in both kinematic limits.

In Section \ref{sec:response} we investigated a possible mechanism for the increased stability. We showed that if the star contracts, the total mass fraction of DM inside the star decreases, so the total mass enclosed by the star also decreases. This added stability is therefore completely agnostic to any particle physics specific to the DM, so long as any consideration of such physics keeps the DM non-relativistic and non-interacting with the baryonic matter. In fact, if the DM possesses some self-annihilation cross-section, like in the model discussed in \citet{ds}, this may further bolster the stability by providing an extra source of energy and entropy.

This brings up a potential limitation of our stability analysis. Our procedure was preformed under the assumption that the mass, total entropy, and angular momentum were fixed.  If we include additional physics, such as rotation, or DM self interactions that inject entropy into the star, our analysis may fall short. Specifically, in \citet{lai1993}, it is shown that a sufficient but not necessary condition for the onset of instability is that both the first and second derivatives of the total energy vanish for the case that total energies parameterized by multiple quantities. However, our analysis involves a sequence of configurations with only one parameter, the central density $\rho_c$, so this condition is sufficient \emph{and} necessary. 

In Section \ref{sec:laneemdenf}, we have determined the equilibrium stellar density profile for an SMS embedded in a field of DM at a constant density. We note that the profile closely follows that of a standard $n=3$ polytrope until the two densities become comparable, whereupon they radically diverge to form a slightly smaller total configuration. The standard and new profiles are quite similar for low DM mass fractions, as expected. Using these new profiles, the change in critical central density is determined quantitatively by integrating them to calculate the structure coefficients that directly determine the critical central density. We find that for all SMS masses and DM mass fractions considered, considering the new equilibrium density profile would serve to slightly lower the increased critical central density due to the DM presence alone.

Finally, in Section \ref{sec:ppi}, we considered by what degree Hydrogen burning could impact the evolution of these SMSs and how DM influences that burning. We also analyzed how neutrino emission from the high-entropy plasma would take away energy, potentially faster than what is generated from fusion, as the star approaches the critical central density from below. We found that, considering both energy generation from fusion and energy losses from neutrino emission, DM could allow heavier stars than previously predicted to stably burn on the main sequence at some point in their lives, increasing the rough upper limit of ${\sim}30,000\,M_\odot$ to about $50,000\,M_\odot$, depending on when one considers nuclear burning to be ``important'' to the evolution of the star.

More investigation is needed to understand these extreme objects. If they are to explain the high mass SMBHs in the early universe by providing a supermassive seed black hole, there should be enough of them formed to account for all or most of them. This is complicated by a suspect formation history, since if the collapsing gas is cool enough to contain molecular Hydrogen instead of atomic Hydrogen, the enhanced cooling efficiency via rotational and vibrational modes \cite{1983ApJ...271..632P} could favor a fragmented formation into many stars instead of one SMS. However, enhanced cooling from early molecular Hydrogen formation via Beyond the Standard Model (BSM) channels (e.g.\ in \citet{PhysRevLett.96.091301}) can lead to very early star formation. Whether SMS formation may accompany this early star formation is an open question. Additionally, if a sufficient population of primordial black holes (PBHs) formed with low enough mass to fully evaporate before $z\sim 10$, the extra heating provided can suppress molecular Hydrogen formation \cite{PhysRevD.109.123016}, possibly allowing another channel to form SMSs. A similar result can also be reached by considering the decay of relic particles \cite{lu2024directcollapsesupermassiveblack}.


We would expect that nuclear burning, i.e.\ fusing bare nucleons into larger composite nuclei, would lower the critical central density for the onset of collapse since the mean molecular weight increases. The gas pressure is provided primarily by non-relativistic baryons, which is proportional to the total number of particles, so nuclear fusion reduces this number and hence lowers the gas pressure if sufficient burning takes place. While sufficient burning would serve to slow the SMS's contraction by injecting energy in the region surrounding the core, the critical central density is being reduced nonetheless, hastening the approach to instability if the chemical composition of the star is significantly changed.


Including DM further complicates this picture. We saw that sufficient DM will change the evolution of some SMS masses that would normally collapse via GR instability, making them also vulnerable to burning induced instability. As DM raises the critical central density, it also permits these stars to host greater central temperatures, making neutrino emission more efficient at carrying away entropy. This would cause the SMS to collapse on a different adiabat, specifically one at lower entropy than expected, potentially altering the prospects for post-collapse detection.

If such stars existed and underwent sufficient burning, they could be detected from their collapse directly if they then explode (instead of directly onto a black hole), and indirectly from enriching their surrounding medium with burning products both pre and post-collapse~\cite{woosley1977neutrino, Fuller_1997}. The enhanced neutrino emission from DM would also alter the gravitational wave signature from an anisotropic neutrino pulse as the homologous core's entropy would be reduced for a given mass (see e.g.~\citet{lifuller} for details on this neutrino driven gravitational radiation for standard SMSs).

The authors would also caution those who may be unfazed by the supposedly `modest' amount of DM invoked in this study. Consider a $10^6\,M_\odot$ SMS with ${\sim}1\%$ DM by mass. If the DM is at a constant density within the star, such a DM mass fraction corresponds to a density of ${\sim}10^{20}\,\mathrm{GeV\,cm^{-3}}$, which would be an over-dense region of DM about 20 orders of magnitude more dense than the expected background at redshift $z\sim10$. 

If these DM supported SMSs had existed, this huge accumulation and concentration of DM greatly constrains models for DM particles. Specifically, models where these particles cannot alter their phase space density and clump together through self and/or standard model interactions would clearly be disfavored in this regard. An example of a model where particles can alter their phase space density appropriately is discussed in \citet{Feng_2021}, which in turn also provides a different mechanism for seeding SMBHs. Densities approaching the requisite amount of DM needed to stabilize SMSs through purely gravitation can be reached via adiabatic contraction of a baryonic cloud in a DM halo (e.g.~\citet{1986ApJ...301...27B}), but would still need an additional source of heating to substantially change the stability of the SMS (e.g.~\citet{Freese_2009} for the case of DM self-annihilation).

Moreover, the huge concentration of DM that we discuss here likely also constrains the manner and environments of SMS formation. Whether such environments can be found, and whether a component of the DM can be compatible with the concentrations discussed here, remain intriguing and open questions. In the end, such speculation may be warranted given our relative ignorance of DM and dark sector physics, and given the existence of SMBHs at very high redshift.

Here we have used a very simplistic model, a non-rotating and isolated SMS, to explore the physical effects of a large DM content. Nonetheless, our findings pose several open questions. First, with the inclusion of DM, what is the interplay of nuclear burning, hydrodynamics, energy transport, neutrino emission, etc., in the evolution of the SMS up to instability and the subsequent collapse to a black hole? Second, are there specific observational ``handles'' on the collapse of a SMS with or without DM? For example, is the gravitational radiation signature \cite{lifuller} different in these two cases, and to what extent would it depend on the amount of DM? Third, in this paper we have not considered the effects of DM on scenarios where a supermassive core undergoes rapid accretion, as in the \emph{hylotropic} configurations proposed in Ref.~\onlinecite{10.1111/j.1365-2966.2009.15916.x}. The stabilizing effect of DM on such hylotropic configurations has been recently considered~\cite{haemmerle}. These scenarios may provide the most plausible route for the production of an SMBH from an SMS-like object.

Finally, perhaps the most significant issue is how an SMS would form with the huge quantities of DM considered in this work and in others that invoke DM effects in the production of SMBHs. However, there is urgency for consideration of these issues because of the existence of SMBHs at very early epochs in the history of the Universe, and because both the sources and properties of DM remain mysterious.


\begin{acknowledgments}
We would like to acknowledge fruitful discussions with Katie Freese and Thomas Wong. This work was supported in part by National Science Foundation (NSF) Grant  No.\ PHY-2209578 at UCSD and the {\it Network for Neutrinos, Nuclear Astrophysics, and Symmetries} (N3AS) NSF Physics Frontier Center, NSF Grant No.\ PHY-2020275, and the Heising-Simons Foundation (2017-228). 
\end{acknowledgments}

\bibliography{DMSMS}


\end{document}